# Robust Kalman filter-based dynamic state estimation of natural gas pipeline networks

Liang Chen [a], Peng Jin [b], Jing Yang [b], Yang Li [c,*], Yi Song [d]

[a] *School of Automation, Nanjing University of Information Science and Technology, Nanjing, 210044, China*

[b] *State Grid Customer Service Center, Tianjin, 300309, China*

[c] *School of Electrical Engineering, Northeast Electric Power University, Jilin, 132012, China*

[d] *StateGrid Economic and Technological Research Institute Co., Ltd, Beijing, 102209, China*

ABSTRACT

To obtain the accurate transient states of the big scale natural gas pipeline networks under the bad data and non-zero mean noises conditions, a robust Kalman filter-based dynamic state estimation method is proposed using the linearized gas pipeline transient flow equations in this paper. Firstly, the dynamic state estimation model is built. Since the gas pipeline transient flow equations are less than the states, the boundary conditions are used as supplementary constraints to predict the transient states. To increase the measurement redundancy, the zero mass flow rate constraints at the sink nodes are taken as virtual measurements. Secondly, to ensure the stability under bad data condition, the robust Kalman filter algorithm is proposed by introducing a time-varying scalar matrix to regulate the measurement error variances correctly according to the innovation vector at every time step. At last, the proposed method is applied to a 30-node gas pipeline networks in several kinds of measurement conditions. The simulation shows that the proposed robust dynamic state estimation can decrease the effects of bad data and achieve better estimating results.

## 1. INTRODUCTION

In comparison with the traditional coal-fired power units, gas-fired electric generators can response to the power load fluctuation rapidly, enhancing the operating flexibility of electrical energy systems [1~3]. This may help to improve security of power system with large scale renewable energy. The random change of the natural gas consumptions due to the uncertainties of renewable energies makes it essential for obtaining the accurate dynamic states just like the pressures and mass flow rates of the natural gas pipeline networks to ensure the security and optimal operation of the integrated energy system containing electric powers and natural gases [4~7]. To capture the states, the pipeline networks have to be equipped with a mount of measuring devices, which requires heavy investments. Even so, it is impossible to install sensors at every node of the networks and obtain all states. On the other hand, the measuring devices experience random errors and bad data unavoidably, so the measured data cannot be applied directly to the leak detections [8,9] and control problems [10] before state estimations.

In recent years, some research works about state estimations for natural gas pipeline networks have appeared [11~15]. These works are based on the nonlinear partial differential equation (PDE) describing the characteristics of transient gas flow [16,17]. To linearize the PDEs of gas systems, in [18,19], the dynamic states are redefined as deviations from steady states for linearization purpose. In [20], an iterative linear approximation is proposed, which is used to solve optimal power flow problem, rather than the linearization of PDEs. In [11], a state estimation method for natural gas pipe lines by using the linearization of PDEs [20-22] is proposed. To improve the estimating performance, the state estimation with a pair of Kalman filter-based estimators running in parallel is carried out. Some researchers establish the state estimation model applying the PDEs directly, and solve the model by nonlinear algorithms. In [12], the extended Kalman filter (EKF) is chosen to design an efficient observer for natural gas transmission system, and then an algorithm is proposed to handle the discontinuities that appear in the dynamic model of a gas transmission networks. In [13], a two-step Lax-Wendroff method is used for the discretization of the PDEs to obtain finite-dimensional discrete-time state-space representations, and the particle filter that fits for the nonlinear filter problems is applied to estimate the transient states. These methods can capture the accurate states in the transient processes. However, as the scale expands, the centralized implementation of the Kalman filter has severe limitations such as tuning, scalability, unacceptable calculating loads and lack of robustness in the case of sensor failures. Aiming at this problem, a strategy for the distributed and decentralized state estimation of state variables in big scale systems is proposed in [14]. In addition, the algorithm for a joint state and parameter estimation problem for large scale networks of pipelines is presented in [15], and the gradient descent algorithm is applied to solve the optimization problem.

---

∗Corresponding author.

*E-mail address*: liyang@neepu.edu.cn (Yang Li).

In practical systems, the supervisory control and data administration system experience random errors and bad data inevitably due to the sensor error and electromagnetic interference. The existing state estimation methods based on Kalman filter can reduce the random errors to some extent, but is vulnerable to bad data. To solve this problem, a variety of improved Kalman filter algorithms are proposed. Based on variational Bayesian technique, an adaptive Kalman filter for linear Gaussian state-space models is proposed in [23,24], which has better robustness to resist the uncertainties of process and measurement noise covariance matrices, as well as the colored measurement noise. In [25], a robust filter in a batch-mode regression form is developed to process the observations and predictions together, making it very effective in suppressing multiple outliers. To ensure the stability of the unscented Kalman filter, the unknown time-varying matrix is introduced to describe the prediction error of the unscented transformation in [26]. It is can be seen that the robustness of the Kalman filter is a big challenge for its practical applications, attacking more and more researchers' attentions.

The existing gas pipeline network state estimation methods based on Kalman filter solve the transient flow equations by using a numerical approximation technique called finite element methods. These estimation methods have the following drawbacks:

1) The gas pipe lines are divided into several linear elements, and the number of variables increases accordingly. For large scale networks with long pipes, to ensure the estimating accuracy, many linear elements have to be created, as well as the states, causing the low measurement redundancy and heavy computation load.
2) The algorithm robustness against bad data is not considered. Once the measurements in the practical system experience bad data or non-zero mean noises, the estimating performances cannot be guaranteed.

To cope with the above problems, and obtain the accurate transient states of large scale gas pipeline networks under practical operating conditions, this paper focuses on the robust dynamic state estimation method along with the following contributions:

1) To deal with the problem that the transient equations are less than the states, the boundary conditions are used as supplementary constraints, and the linear process functions are established. The zero mass flow rates at the sink nodes are taken as virtual measurements, improving the measurement redundancy.
2) A time-varying scalar matrix is proposed to regulate the measurement variance matrix according to the innovations, making that the measurements can correct the predicted states accurately, and then increasing the robustness of the Kalman filter against the bad data and non-zero mean noises.
3) The proposed dynamic state estimation method is tested on a 30-node natural gas pipeline network under the normal measurement condition, bad data condition and non-zero mean noises condition. The simulation results show that the proposed approach manages to perform robust DSE of natural gas pipeline networks and the performance of the proposed method is better than the traditional Kalman filter.

The rest of this paper is organized as follows. Section 2 builds the mathematical model of the DSEs for gas pipelines. Section 3 proposes the robust Kalman filter algorithm based on the time-varying scalar matrix. Section 4 shows and analyzes the simulation results under various conditions. Finally, Section5 concludes this paper.

## 2. DSE modeling of gas pipeline network

As the basic components of natural gas systems, the pipelines can store a certain amount of natural gases due to the compressibility, which is called the linepack storage. For the natural gas systems with a large number of pipelines, the linepack storage drives the dynamic process of the nodal pressure and mass flow of the natural gas in the pipelines. To describe the above physical dynamic mathematically, the state space representation of transient gas flows is introduced.

*2.1. State space representation of transient gas flow*

Under stable operating conditions, the gas states such as the nodal pressure and mass flow rate in pipelines are constant. However, in practical natural gas systems, the stable operating condition can be disrupted easily due to the continuous changes of gas loads, supplies of gas companies and other operational activities, causing the fluid dynamic process along pipelines, which can be represented by a set of nonlinear partial differential equations (PDE). These PDEs are derived from the momentum conservation principles and materiel balances, which is given as the following state space form [11,4,27]:

$$\frac{\partial \rho}{\partial t} + \frac{\partial (\rho u)}{\partial x} = 0 \tag{1a}$$

$$\frac{\partial (\rho u)}{\partial t} + \frac{\partial (\rho u^2)}{\partial x} + \frac{\partial p}{\partial x} = -\frac{f \rho u |u|}{2d} - \rho g \sin \theta \tag{1b}$$

where, $\rho$, $p$ and $u$ are the density, pressure and gas velocity respectively; $t$ and $x$ are the time and spatial coordinate respectively; $f$ is the friction factor; $g$ is the gravitational constant; $\theta$ is the pipe inclination angle; $d$ is the pipe diameter. The second term on the left-hand side of (1b) describes the convective effect of the natural gas, which can be omitted when the gas velocity is significantly smaller than the sound speed in practical operation [4,11]. Further, we assume that the pipelines are horizontal, hence the inclination angle is zero, and the second term on the right-hand side of (1b) can also be omitted. Under the condition that the gas pressure is less than 10 bar and the temperature is lower than 20 °C, the functions between the pressure, mass flow

rate and the gas density are as follows [11,4]:

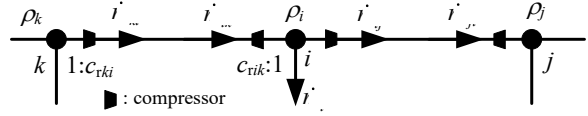

Fig. 1. The general model of natural gas pipelines.

$$p = c^2 \rho \tag{2}$$

$$\dot{\iota} \tag{3}$$

where, $c^2 = ZRT$; $\dot{\iota}$ and $a$ are the mass flow rate and cross section area of pipelines respectively; $Z$, $R$ and $T$ are the gas compressibility factor, the specific gas constant and the gas temperature, respectively. For the natural gas, $c^2$ is constant if the temperature is not changing. Here, the absolute value of average gas velocity $|\bar{u}|$ is used instead of $|u|$. Under the aforementioned assumptions, equation (1) can be rewritten as the following equation [4]:

$$\frac{\partial \rho}{\partial t} + \frac{\partial \dot{\iota}}{a \partial x} = 0 \tag{4a}$$

$$\frac{\partial \dot{\iota}}{a \partial t} + c \frac{\partial}{\partial x} + \frac{\cdot}{2da} = 0 \tag{4b}$$

The simplified partial differential equation (4) is able to describe the practical natural gas dynamic process in continuous form, which should be transformed to difference equation before numerical solving.

*2.2. Discretization of PDE*

A generalized model of natural gas pipelines is shown in Fig. 1. Assuming that the node numbers of the three nodes of pipelines are $i$, $j$ and $k$, respectively, and $k<i<j$. The gas flow direction is defined as from the smaller node to the bigger node. Thus, the gas flow directions of pipeline $k$-$i$ and $i$-$j$ are $k \rightarrow i$ and $i \rightarrow j$, respectively. The gas densities are $\rho_k$, $\rho_i$ and $\rho_j$, while the mass flow rates on the two sides of the pipeline $i$-$j$ are $\dot{\iota}$ and $\dot{\iota}$, respectively. The mass flow rate of gas load at node $i$ is $\dot{\iota}$. For simplification, the compressor at $k$ end of pipeline $ki$ is modeled as a constant ratio of densities $c_{rki}$. The positive direction is defined as flowing out, thus the mass flow rate values are negative for the source nodes. Every natural gas pipeline can be modeled according to Fig. 1, and satisfy the PDEs (4). One of the common methods for solving PDEs is the Euler finite difference technique, which is applied to solve (4) in our works. The differencing scheme is [28]:

$$\frac{\partial X}{\partial t} \approx \frac{1}{2}\left(\frac{X_{s+1,t+1} - X_{s+1,t}}{\Delta t} + \frac{X_{s,t+1} - X_{s,t}}{\Delta t}\right) \tag{5a}$$

$$\frac{\partial X}{\partial x} \approx \frac{1}{2}\left(\frac{X_{s+1,t+1} - X_{s,t+1}}{\Delta x} + \frac{X_{s+1,t} - X_{s,t}}{\Delta x}\right) \tag{5b}$$

$$X \approx \frac{1}{4}\left(X_{s+1,t+1} + X_{s,t+1} + X_{s+1,t} + X_{s,t}\right) \tag{5c}$$

where, $X$ represents the generalized stats; $\Delta t$ and $\Delta x$ are the time and spatial step widths respectively; the subscripts $s$ and $t$ represent the $s$th node and time step $t$, respectively. Equation (4) can be transformed to (6) according to the differencing scheme (5) for pipeline $i$-$j$.

$$c_{rji}\rho_{j,t+1} - c_{rji}\rho_{j,t} + c_{rij}\rho_{i,t+1} - c_{rij}\rho_{i,t} + \frac{\Delta t}{L_{Pij} a_{ij}}\left(\dot{\iota} \cdots \right) \tag{6a}$$

$$\dot{\iota} \cdots \frac{a_{ij}\Delta t c^2}{L_{Pij}}\left(\cdots\right) \cdots \frac{|\bar{v}|\Delta t}{4 d_{ij} a_{ij}}\left(\cdots\right) \tag{6b}$$

where, the subscript $ij$ represents the pipeline $i$-$j$; $L_{ij}$ is the length of pipeline $i$-$j$. In addition to the difference equations, some boundary conditions should be satisfied.

*2.3. Boundary conditions*

The gas flow dynamic processes and gas density distribution are influenced by the changes of operating conditions at both the source and sink nodes. Under the practical operating condition, the mass flow rates at sink nodes are changing, while the gas density is constant at the source nodes. These constraints are defined as boundary conditions.

The boundary condition of the mass flow rate balance at sink nodes is represented as:

$$\sum_{k\in i, k<i} \dot{m}_{\cdots} \quad \overline{\phantom{xxx}}_{j\in i, j>i} \quad \cdots, \quad i \in \qquad (7)$$

where, $k \in i$ means the node $k$ connected to node $i$ with pipeline $i$-$k$; $i \in$ represents $i$ is the source node. The first and second terms of (7) represent the sum of mass flow rates from node $i$ and to node $i$, respectively.

For the source nodes, the assumption is made that the gas capacities are big enough so that the gas density is able to maintain constant in the dynamic process. This forms the following boundary condition:

$$\rho_{i,t} = \rho_{Si,0}, \quad i \in N_{\text{Source}} \qquad (8)$$

where, $\rho_{Si,0}$ is the gas density at source node $i$ at initial time; $i \in N_{\text{Source}}$ represents $i$ is the source node.

The boundary conditions supply more constraints, making it possible to build the mathematical model of DSE for the natural gas pipeline networks.

### 2.4. Mathematical model of DSE

The DSE based on Kalman filter includes two basic steps, the prediction and filtering. In the prediction step, the difference equations and boundary conditions are used to predict the states. In this work, the gas densities at nodes and mass flow rates at the two ends of pipelines are taken as states. Hence, for a natural gas pipeline network with $n_N$ nodes and $n_L$ pipelines, the number of states is $n_N + 2n_L$. The state vector at time step $t$ is $x_t = [x_{r,t}, x_{m,t}]^T$, $x_{r,t} = \qquad \cdots \qquad$, $x_{m,t} = \cdots$, $i<j$. $x_{r,t}$ is the state vector of gas densities; $x_{m,t}$ is the state vector of mass flow rates, and the dimension of $x_{m,t}$ is $2n_L$. Based on the above state definition, (6) can be rewritten as

$$\begin{bmatrix} A & A \\ A & A \end{bmatrix} \begin{bmatrix} x \\ x \end{bmatrix} \quad \begin{bmatrix} A & A \\ A & A \end{bmatrix} \begin{bmatrix} x \\ x \end{bmatrix} \qquad (9)$$

$$A_{11}(l,i) = \qquad \in \qquad (10)$$

$$A_{12} = \begin{bmatrix} & & 0 \\ & \ddots & \\ 0 & & \end{bmatrix} \quad A \quad \mathbb{R} \qquad (11)$$

$$\alpha = \frac{}{L_{ij} a_{ij}} \propto \propto \qquad (12)$$

$$A_{21}(l,i) = \begin{cases} \propto \propto < \\ \propto \propto > \\ \text{se} \end{cases} \in \qquad (13)$$

$$\beta_l = \frac{a_{ij} \Delta t c^2}{L_{ij}}, i \propto l, j \propto l \qquad (14)$$

$$A_{22}(l,i) = \begin{cases} \times \propto < \\ \propto > \end{cases} \in \qquad (15)$$

$$\gamma_l = \frac{f|\bar{u}|\Delta t}{4 d_{ij} a_{ij}}, i \propto l, j \propto l \qquad (16)$$

where, $(l, i)$ represents the row $l$ and column $i$; $i \propto$ means node $i$ is one end of pipeline $l$. For the natural gas pipeline systems, the total number of state variables is $n_N + 2n_L$, but the number of equations in (9) is $2n_L$. Equation (9) cannot be solved because the states are more than the equations. The boundary conditions can be applied as the supplementary constraints. Similarly, (7) and (8) can be rewritten in the matrix form.

$$\begin{bmatrix} B & 0 \\ 0 & B \end{bmatrix} \begin{bmatrix} x \\ x \end{bmatrix} \begin{bmatrix} u \\ u \end{bmatrix} \tag{17}$$

$$B_{11}(s,i) = \quad \in \tag{18}$$

$$B_{22}(j,2l) = \quad \in \tag{19a}$$

$$B_{22}(i,2l-\ =\ \quad \propto \tag{19b}$$

where, $u_{r,t+1} = [\ldots, \rho_{Si,0}, \ldots]^T$, $i \in$ , $u_{r,t+} \in$ ; $u_{m,t+1} = [\ldots, \dot{i}_+, \ldots]^T, i \in$ , $u_{m,t+} \in$ ; $n_{Source}$ and $n_{Sink}$ are the numbers of source and sink nodes, respectively; $s$ is the source nodes number, $s = 1, 2, \ldots, n_{Source}$; $\mathbf{0}$ is the zero matrix and the subscript of $\mathbf{0}$ is the dimension.

Equation (9), (17) can be written as the following concentrated form:

$$\mathcal{A}x_{t+1} = \mathcal{B}x_t + \mathcal{U}_{t+1} \tag{20}$$

where, $\mathcal{A} = \begin{bmatrix} A & A \\ A & A \\ B & 0 \\ 0 & B \end{bmatrix}$, $\mathcal{B} = \begin{bmatrix} A & A \\ A & A \\ 0 & 0 \\ 0 & 0 \end{bmatrix}$, $\mathcal{U}_{t+1} = \begin{bmatrix} 0 \\ 0 \\ u \\ u \end{bmatrix}$.

Premultiplied by the inverse matrix of the left matrix, (20) becomes

$$x_{t+1} = Fx_t + u_{t+1} \tag{21}$$

where, $F = \mathcal{A}^{-1}\mathcal{B}$, $u_{t+1} = \mathcal{A}^{-1}\mathcal{U}_{t+1}$.

Equation (21) is the system equation of the gas pipeline networks, which is used to represent the dynamic processes. Additionally, the measured pressure and mass flow rate information from supervisory control and data administration systems provide redundant constraints for the DSE. In this work, assume that all of the nodes are equipped with flow meters and barometers, hence such information is taken as the measurement vectors. The measurement function is

$$z_{t+1} = H_{t+1}x_{t+1} \tag{22}$$

$$H_{t+} = \begin{bmatrix} I & 0 \\ 0 & H \end{bmatrix} \tag{23}$$

where, $H' \quad = \quad H \quad - \quad = \quad H \in \mathbb{R}$ $z_{t+1} = [z_{r1,t+1}, z_{r2,t+1}, \ldots, z_{m_N,t+}, z_{m1,t+1}, z_{m2,t+1}, \ldots, z_{mn_N,t+}]^T$; $I$ is an identity matrix; $z_{ri,t+1}$ and $z_{mi,t+1}$ are the note pressure and mass flow rate measurements at time instant $t+1$ respectively.

Up to now, the DSE model of gas pipeline networks is formed

$$\begin{cases} \dot{x} & Fx + u + v \\ z & Hx + w \end{cases} \tag{24}$$

where, $v_t$ and $w_{k+1}$ are the system and measurement error vectors respectively, $v_t \in \mathbb{R}$ , $w_{t+} \in \mathbb{R}$ . The variance matrixes of $v_{t+1}$ and $w_{k+1}$ are $Q_{t+1}$ and $R_{t+1}$ respectively. In this paper, the errors are assumed to satisfy the normal distribution. However, the practical measurement experience bad data inevitably due to the electromagnetic interference or transmission errors. Additionally, in some cases, the mean values of the measurement errors are not zero, so the errors are non-zero mean noises.

The states in (24) should be estimated by the DSE algorithm based on Kalman filter.

## 3. Robust dynamic state estimation algorithm

One of the commonly used dynamic state estimation algorithms is Kalman filter, which is used in this work.

### 3.1. Kalman filter

Kalman filter algorithm [29] includes two basic steps: prediction and filtering.

**Prediction step.** Given the initial estimated state $\hat{\ }_0$ and its covariance matrix $P_{t|t}$, the state prediction $\tilde{\ }$ and predicting covariance matrix $P_{t+1|t}$ can be calculated by (25), (26).

$$\tilde{x}_{t+1|t} = F\hat{x}_{t|t} \quad (25)$$
$$P_{t+1|t} = FP_{t|t}F^{\mathrm{T}} + Q_t \quad (26)$$

**Filtering step.** The predicted state $\tilde{x}_{t+1|t}$ should be corrected, and the estimated state $\hat{x}_{t+1|t+1}$ and covariance matrix $P_{t+1|t+1}$ are obtained.

$$K_{k+1} = P_{t+1|t}H^{\mathrm{T}}(HP_{t+1|t}H^{\mathrm{T}} + R_{t+1})^{-1} \quad (27)$$
$$\hat{x}_{t+1|t+1} = \tilde{x}_{t+1|t} + K_{k+1}(z_{t+1} - H\tilde{x}_{t+1|t}) \quad (28)$$
$$P_{t+1|t+1} = P_{t+1|t} - K_{k+1}HP_{t+1|t} \quad (29)$$

The right part of (27), $P_{e,t+1} = HP_{t+1|t}H^{\mathrm{T}} + R_{t+1}$, is the innovation covariance matrix, which represents the errors between measurements and their predicting values. When the measurements experience bad data, $P_{e,t+1}$ cannot represent the real errors correctly. Thus, the estimating performance of Kalman filter decreases. In this work, we introduce a time-varying scalar to regulate the measurement variance matrix, making Kalman filter robust.

*3.2. Time-varying scalar*

Kalman filter estimates the states by trading off the predicted states against the measurements according to the measurement and prediction variances. If the measurements experience bad data, Kalman filter would not correct the predicted states accurately, and result in a poor estimation results. Aiming at this problem, to make the algorithm robust to the bad data, a time-varying scalar matrix $\mu_t$ is proposed to regulate the measurement variance matrix. The objective of $\mu_t$ is to fulfill (30).

$$P_{e,t+1} = HP_{t+1|t}H + \mu_{t+1}R_{t+1} \quad (30)$$

To obtain the value of $\mu_{t+1}$, the sliding window method is used to estimate the innovation covariance matrix:

$$P_{e,t+1} = \frac{1}{m_{\mathrm{W}}}\sum_{i=0}^{m_{\mathrm{W}}-1} e_{t+1-i}e_{t+1-i}^T = HP_{t+1|t}H + \mu_{t+1}R_{t+1} \quad (31)$$

where, $e_{t+1} = z_{t+1} - H\tilde{x}_{t+1|t}$ is the innovation vector; $m_{\mathrm{W}}$ is the window length. By solving (31), $\mu_{t+1}$ can be obtained as

$$\mu_{t+1} = \left(\frac{1}{m_{\mathrm{w}}}\sum_{i=0}^{m_{\mathrm{w}}-1} e_{t+1-i}e_{t+1-i}^T - HP_{t+1|t}H\right)R_{t+1}^{-1} \quad (32)$$

The obtained $\mu_{t+1}$ by (32) may be non-diagonal, as a result, the inversion of the innovation covariance matrix is singular probably. Aiming at this problem, a diagonal matrix $\mu'_t$ is defined

$$\mu'_{t+1} = \mathrm{diag}(\mu'_1, \mu'_2, \cdots, \mu'_{2n_N}) \quad (33)$$

where, $\mu'_i = \mu_{t+1,ii}$, $i = 1, 2, \ldots, 2n_N$, $\mu_{t+1,ii}$ is the $i$th diagonal element. Equation (27) becomes

$$K_{k+1} = P_{t+1|t}H^{\mathrm{T}}(HP_{t+1|t}H^{\mathrm{T}} + \mu'_{t+1}R_{t+1})^{-1} \quad (34)$$

With the help of the scalar, the modified robust Kalman filter can decrease the influence of bad data, and a better performance can be obtained than the traditional Kalman filter.

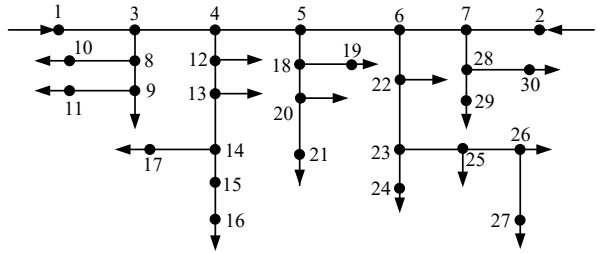

Fig. 2. The 30-node test system. The tiny black points are pipeline conjunction nodes. Node 1 and 2 are the inlet nodes, the others are outlet nodes. The outlet nods without arrows means that the gas loads are zero.

## 4. Case study

In this section, the proposed DSE method is applied to a 30-node natural gas pipeline network shown in Fig. 2. The time-varying mass flow rates of gas loads are simulated artificially and the dynamic processes of the node pressures and mass flow rates are calculated by (6)~(8). The simulation results are taken as true values, and the measurements are derived by adding random numbers to the real values. The traditional Kalman filter and the proposed algorithm are used to estimate the states of the gas pipeline system. The DSE is carried out in the following three measurement conditions: normal condition, bad data condition and non-zero mean noises condition.

*4.1. Description of the test system*

The test system includes 30 nodes and 29 pipelines, which is used to evaluate the performances of DSE methods. The lengths and cross-sectional diameters are given in Tab.1. The friction factor $f$ and the gas speed $c^2$ are 0.015 and 340 m/s respectively. Gas is supplied at the source node 1 and 2, and the pressures of these two nodes are 27.8 bar and 28.5 bar respectively. We assume that the capacities of gas sources are infinite, which means the pressures are constant in the dynamic processes.

Table 1
Parameters of gas pipelines.

| Pipeline | Length (km) | Diameter (m) | Pipeline | Length (km) | Diameter (m) | Pipeline | Length (km) | Diameter (m) | Pipeline | Length (km) | Diameter (m) |
|---|---|---|---|---|---|---|---|---|---|---|---|
| 1,3 | 5 | 0.6 | 8,10 | 7 | 0.2 | 5,18 | 10 | 0.4 | 25,26 | 9 | 0.2 |
| 3,4 | 3 | 0.6 | 9,11 | 5 | 0.4 | 18,20 | 3 | 0.2 | 26,27 | 4 | 0.2 |
| 4,5 | 4 | 0.5 | 4,12 | 4 | 0.4 | 20,21 | 7 | 0.2 | 7,28 | 2 | 0.2 |
| 5,6 | 6 | 0.5 | 12,13 | 8 | 0.4 | 18,19 | 2 | 0.2 | 28,29 | 7 | 0.2 |
| 6,7 | 7 | 0.5 | 13,14 | 10 | 0.4 | 6,22 | 10 | 0.4 | 28,30 | 5 | 0.2 |
| 2,7 | 2 | 0.5 | 14,15 | 9 | 0.2 | 22,23 | 6 | 0.2 | | | |
| 3,8 | 3 | 0.4 | 15,16 | 10 | 0.2 | 23,24 | 7 | 0.2 | | | |
| 8,9 | 5 | 0.2 | 14,17 | 4 | 0.2 | 23,25 | 4 | 0.2 | | | |

The simulation interval is 15 minutes over the time horizon of 24 hours. All nodes are equipped with flow meters and barometers, so the pressures and mass flow rates can be measured. The standard deviations of pressure and mass flow rate measurements are 0.01 bar and 2% respectively. Note that the mass flow rates at the conjunction nodes are zero, which can be taken as virtual measurements. The errors of virtual measurements are smaller than others, and the errors are set to 0.001 in our work.

*4.2. Normal measurement condition*

In this section, the DSE is carried out under the normal measurement condition that the measurement errors obey the Gauss distribution with zero mean. The traditional Kalman filter and robust Kalman filter are used to estimate the states, and the results are compared by the filter coefficient $\varepsilon$ [30,31].

$$\varepsilon = \frac{\sum \hat{z}_t - \tilde{z}_t}{\sum z_t^+ - z_t} \qquad (35)$$

where, $\hat{z}_t$ is the estimated value of measurements, which can be calculated by the estimated states; $z_t^+$ is the true value of measurements without errors.

The filter coefficients of the two kinds of DSE methods are shown in Tab.2. Nodes 1 and 2 are source nodes, the pressures of which are constant, so the filter coefficients are not computed. All results in Tab. 2 are smaller than 1, meaning that both the DSE based on Kalman filter and robust Kalman filter are effective. It should be noted that the virtual measurement errors of mass flow rates at the sink nodes are very small and the denominator of (35) is approximately equal to the numerator, as a result, the filter coefficients is almost 1. Besides that, most of the coefficients of the robust Kalman filter are smaller than the traditional Kalman filter, meaning that the performance of the robust Kalman filter is better.

Table 2
Filter coefficients of DSEs

| Node | Pressure | | Mass flow rate | | Node | Pressure | | Mass flow rate | |
|---|---|---|---|---|---|---|---|---|---|
| | KF | RKF | KF | RKF | | KF | RKF | KF | RKF |
| 1 | / | / | 0.1111 | 0.1084 | 16 | 0.0355 | 0.0363 | 0.0405 | 0.0151 |
| 2 | / | / | 0.4831 | 0.4492 | 17 | 0.0308 | 0.0206 | 0.0061 | 0.0021 |
| 3 | 0.0001 | 0.0001 | 0.9888 | 0.9888 | 18 | 0.0006 | 0.0004 | 0.9888 | 0.9888 |
| 4 | 0.0001 | 0.0001 | 0.9888 | 0.9888 | 19 | 0.0331 | 0.0144 | 0.1634 | 0.0956 |
| 5 | 0.0002 | 0.0001 | 0.9888 | 0.9888 | 20 | 0.0388 | 0.0293 | 0.2903 | 0.2361 |
| 6 | 0.0002 | 0.0001 | 0.9888 | 0.9888 | 21 | 0.0601 | 0.0473 | 0.2087 | 0.1523 |
| 7 | 0.0003 | 0.0002 | 0.9888 | 0.9888 | 22 | 0.0004 | 0.0003 | 0.4232 | 0.3063 |
| 8 | 0.0003 | 0.0002 | 0.9888 | 0.9888 | 23 | 0.0064 | 0.0042 | 0.9888 | 0.9888 |
| 9 | 0.0014 | 0.0011 | 0.5619 | 0.5161 | 24 | 0.0372 | 0.0249 | 0.0661 | 0.0281 |
| 10 | 0.0061 | 0.0049 | 0.5101 | 0.3902 | 25 | 0.0361 | 0.0298 | 0.4338 | 0.2521 |
| 11 | 0.0011 | 0.0008 | 0.1394 | 0.0706 | 26 | 0.0501 | 0.0363 | 0.3376 | 0.2185 |
| 12 | 0.0011 | 0.0007 | 0.4301 | 0.3681 | 27 | 0.0687 | 0.0365 | 0.1316 | 0.0748 |
| 13 | 0.0016 | 0.0011 | 0.2984 | 0.1957 | 28 | 0.0193 | 0.0114 | 0.9888 | 0.9888 |
| 14 | 0.0027 | 0.0021 | 0.5032 | 0.4625 | 29 | 0.0551 | 0.0318 | 0.0944 | 0.0606 |
| 15 | 0.0187 | 0.0245 | 0.6488 | 0.6339 | 30 | 0.0543 | 0.0394 | 0.1021 | 0.0661 |

*4.3. Bad data condition*

In practical systems, the measuring equipments experience bad data inevitably [32], which should be considered in DSEs. In this section, based on the normal measurement condition, the bad data are added to the measurement vectors artificially to test the performances of the proposed DSE for gas pipeline networks. Here, the value of the pressure measurements at node 30 is set to 12, 10.7 and 13.8 bar at time 5, 5.25 and 5.5 hour, 13, 15.5 and 23 bar at time 13.25, 13.5 and 13.75 hour, respectively. Meanwhile, the mass flow rate measurements at node 11 are reset to 3, 2.1, 3 and 2.2 kg/s at time 7.5, 7.75, 8 and 8.25 hour, and 3, 2.1 and 1.7 kg/s at time 15.75, 16 and 16.25 hour, respectively.

The estimating results of Kalman filter and robust Kalman filter are shown in Fig. 3, and the time-varying scalars are shown in Fig. 4. It is can be seen that the curve of Kalman filter (the green dotted line) is deviated from the true value (the blue solid line) when the bad data appeared. This is because the measurement error variance matrix does not correspond with the truth in the bad data condition. However, with the help of the time-varying scalar, the curve of robust Kalman filter (the yellow dotted line) and the true value curve are well coincident all the time. The time-varying scalar can regulate the measurement error variance matrix according to the actual errors precisely and the accurate estimating results can be obtained. When the bad data appeared, the values of the scalar increase dramatically, causing the reduction of the corresponding elements in Kalman gain matrix $K_{t+1}$. As a result, the correcting effects of the measurements decrease and the estimated values are little affected by the bad data.

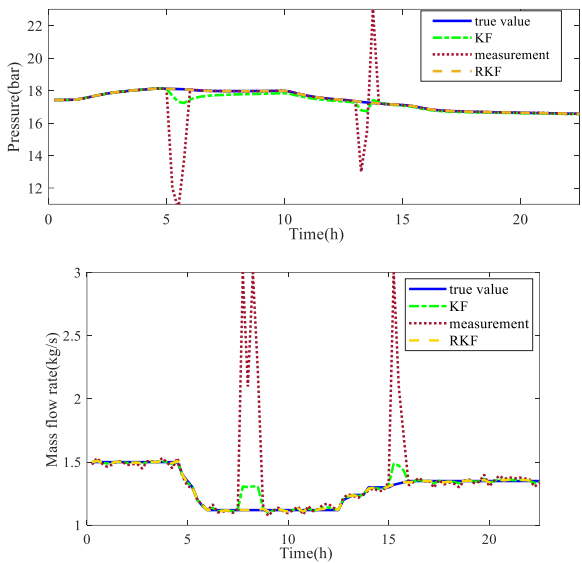
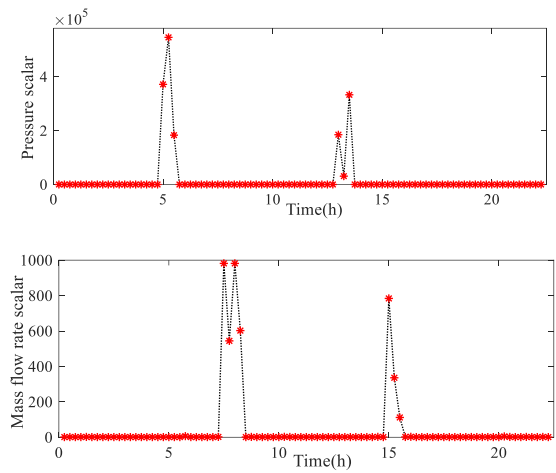

Fig. 3. Estimating results of the pressure and mass flow rate under bad data condition.

Fig. 4. Time-varying scalars of the pressure and mass flow rate under bad data condition.

*4.4. Non-zero mean noises condition*

The errors of measurements in the above two sections are Gaussian white noises, but in practice the non-zero mean noises are more common, which are considered in this section. Based on the condition of section B, we added constant deviations 0.2 bar and 0.1 kg/s to the pressure measurements and mass flow rate measurements from 10~19.75 hour and 5~12.5 hour, respectively.

The estimating results are shown in Fig. 5 and Fig. 6. It can be seen that the estimating curves of Kalman filter deviate from the true value curves obviously under the non-zero mean noises condition. At the same time, we can notice that the scalars increase obviously while the measurement mean values are deviating from zero, and the estimating curves based on robust Kalman filter fit with the true value all the time. Furthermore, the Kalman filter curves deviate from true value gradually at the beginning of the non-zero mean noises. This is because the negative effects of the non-zero mean noises are accumulating continuously throughout the whole process. If the non-zero mean noises exist all the time, eventually, the estimated value of the Kalman filter will coincide with the deviation.

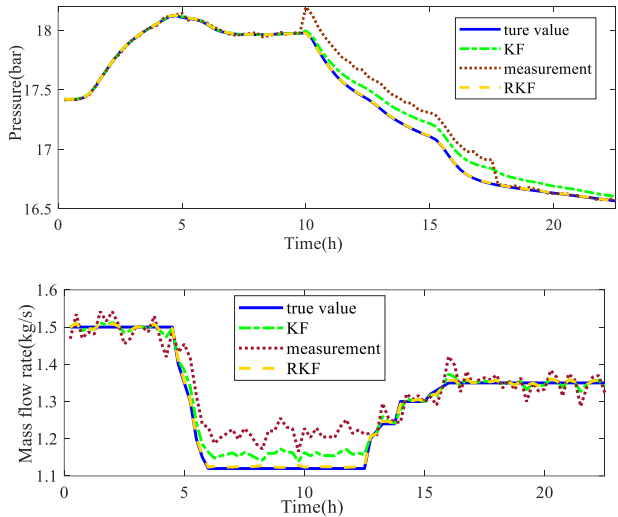
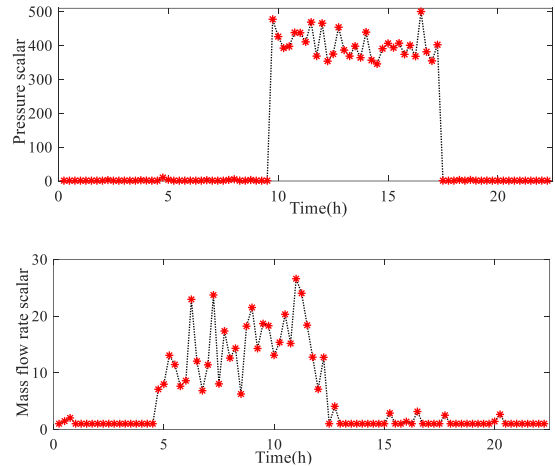

Fig. 5. Estimating results of the pressure and mass flow rate under non-zero mean noises condition.

Fig. 6. Time-varying scalars of the pressure and mass flow rate under non-zero mean noises condition.

## 5. Conclusion

This paper proposes a robust dynamic state estimation method against bad data for natural gas pipeline networks. The method is applied on a 30-node pipeline network in several conditions, and the filter coefficient is used to evaluate the performances of the dynamic state estimation based on the traditional Kalman filter and the proposed robust method. The results show that most of the filter coefficient values of the proposed method are smaller than the traditional Kalman filter in the normal measurement condition. This means that the filtering efficiency of the robust method is better. Furthermore, the two methods are studied under the bad data and non-zero mean measurement noises conditions, and the results show that the proposed method can decrease the effects of bad data, obtaining accurate estimating states under these two conditions.

Future work will focus on the accurate DSE modeling based on the nonlinear transient equations of the pipeline networks with compressors under the operating limitation constraints, and calculating method of predicting error covariance; another interesting topic is the adaptive DSE algorithm that can cope with the model uncertainty due to parameter errors and malicious cyber attacks [33,34]. Besides, it is interesting to extend the presented approach to the DSE of integrated energy systems with integration of new elements such as cogeneration units [35], electrical vehicles [36], modular multilevel converter based high-voltage direct current systems [37] and uncertain renewable generations [38].

### Acknowledgements
This work is partly supported by the Natural Science Foundation of Jilin Province, China under Grant No. 2020122349JC.